\documentclass[review]{elsarticle}

\usepackage{hyperref}
\usepackage{graphicx}%
\newcommand{\be}{\begin{equation}}
\newcommand{\ee}{\end{equation}}
\newcommand{\bea}{\begin{eqnarray}}
\newcommand{\eea}{\end{eqnarray}}

\newcommand{\sn}{{\rm sn}}

\newcommand{\dn}{{\rm dn}}
\newcommand{\cn}{{\rm cn}}

\begin{document}

\begin{frontmatter}
\title{Intrinsic Localized Modes of a Classical Discrete Anisotropic 
Heisenberg Ferromagnetic Spin Chain} 

\author{M. Lakshmanan\corref{mycorrespondingauthor}} 
\cortext[mycorrespondingauthor]{Corresponding author}
\ead{lakshman@cnld.bdu.ac.in}
\author{B. Subash}
\address{Centre for Nonlinear Dynamics, Department of Physics, 
Bharathidasan University, Tiruchirapalli 620 024, India} 

\author{Avadh Saxena}
\address{Theoretical Division and Center for Nonlinear Studies, Los
Alamos National Lab, Los Alamos, NM 87545, USA}

\date{\today} 

%\vskip 5.0cm 
\begin{abstract}
%\vskip 2.0cm 
We report several exact intrinsic localized mode solutions of the classical spin evolution equation of a one-dimensional anisotropic Heisenberg ferromagnetic spin chain in terms of Jacobian elliptic functions. These include one, two and three spin excitations. All these solutions have smooth anticontinuum limits. Their linear stability and semiclassical quantization are also discussed briefly. 
\end{abstract}

\end{frontmatter}

%\linenumbers
%\maketitle 

\section{Introduction} 
Recently there has been considerable interest in identifying and analyzing intrinsic localized modes (ILMs) or discrete breathers \cite{Sieversl,page,flach04} in classical nonlinear Hamiltonian lattices \cite{flach98,aubry94,aubry97}, including magnetic chains \cite{rakh98,zolo,khalack02,sievers97,sievers03,peyrard05} with direct experimental observation \cite{sato04}.  Quantum spin chains have also been studied \cite{tang13}.  ILMs are dynamically localized states, which exhibit periodic oscillations in time and are localized in space.  The standard approach to identify them is to start from an anticontinuum (AC) limit and analytically continue the solution to include discreteness effects. Considerable wealth of information has accumulated on the properties of ILMs over the years in varied types of nonlinear lattices and magnetic chains \cite{kevrekidis,pelinovsky,Noble}.  However, to our knowledge exact analytic solutions of these excitations are very limited in these models. Our main objective in this paper is to obtain a class of exact analytic solutions of ILMs of a classical one-dimensional discrete anisotropic Heisenberg ferromagnetic chain including external magnetic fields, which one can then search experimentally \cite{exp}.  The classical anisotropic spin 
systems are of significant physical interest, e.g. in the context of spin waves in magnetic bodies of arbitrary shape \cite{rivkin05} and spin-torque 
in ferromagnetic layers with spin currents \cite{li04}.  Dynamical spin-spin correlations in the presence of a magnetic field have also been studied in such systems \cite{davis05}.  Materials realization of classical Heisenberg spin chains include CsNiF$_3$ and tetramethyl ammonium manganese trichloride (TMMC). 

Dynamics of anisotropic Heisenberg ferromagnetic spin system with or without magnetic field is of considerable relevance in magnetism, condensed matter physics, materials science, spintronics, statistical physics, nonlinear dynamics, etc. \cite{Laksh11,mertens}. Particularly for large value of spins, an adequate description is to treat the spin angular momentum as classical unit vectors and write down the equation of motion using appropriate spin Poisson brackets \cite{discrete1}. It is also of interest to note that in the continuum limit in one spatial dimension several cases are completely integrable soliton systems such as the pure isotropic case \cite{laksh}, the uniaxial anisotropic case \cite{naka82} and the biaxial anisotropic case \cite{sklyanin}, while no case of the discrete Heisenberg spin system is known to be integrable (although a related version, namely Ishimori chain \cite{ishimori} is completely integrable). However, a number of special solutions in the discrete spin chain case does exist under different situations, as two of us have shown recently \cite{LS}.

In this connection ILMs have been explored in anisotropic spin chains starting from the anticontinuum limit to the nonzero exchange constant case analytically and numerically by several authors \cite{zolo,khalack02,sievers97,speight,nag} and the existence of ILMs under specific interactions has been established. However, to our knowledge, no exact ILM solutions for the spin chains have been reported in the literature. In this paper, we show explicitly that one, two and three (as well as higher) spin excitations can be obtained in terms of Jacobi ellliptic functions. These solutions have appropriate anticontinuum limits. Their linear stability can also be explored systematically. In addition, their semiclassical quantization can be carried out straightforwardly.

The plan of the paper is as follows. In Sec. \ref{model}, we discuss the form of the anisotropic Heisenberg spin chain in the presence of an external magnetic field and present the associated evolution equation for the spin vectors. In Sec. \ref{ilm}, we obtain explicit one-spin, two-spin and three-spin excitations in terms of Jacobi elliptic functions. Section \ref{stability} deals with the linear stability analysis of the above solutions. In Sec. \ref{semiclass}, we present a brief discussion on the semiclassical quantization of the above ILMs. Finally in Sec. \ref {conclusion}, we summarize our main results and present our conclusions.

\section{The Heisenberg Anisotropic Spin Chain: Model and Equation of 
Motion} 
\label{model}

To begin with we briefly introduce the evolution equation for the spins of a one dimensional anisotropic Heisenberg ferromagnetic spin chain modeled by the Hamiltonian \cite{LS}
\be
\label{hamil} 
H=-\sum_{\{n\}}^N(AS_n^xS_{n+1}^x+BS_n^yS_{n+1}^y+CS_n^zS_{n+1}^z) 
-D\sum_n(S_n^z)^2-\vec{\cal H}\cdot\sum_n{\vec{S}}_n , 
\ee 
where the spin components $\vec S_n=(S_n^x,S_n^y,S_n^z)$, are classical unit vectors satisfying the constraints  
\be 
\label{constt}
(S_n^x)^2+(S_n^y)^2+(S_n^z)^2=1, ~n=1,2,...,N. 
\ee 
In Eq. (\ref{hamil}) the sum is over the nearest neighbors, $A$, $B$ and $C$ are the (exchange) anisotropy parameters, $D$ is the onsite anisotropy parameter and $\vec{\cal H}=({\cal H},0,0)$ is the external magnetic field along the $x$ direction (for convenience). 
One can introduce appropriate spin Poisson bracket relations  \cite{discrete1} 
%\be \{S_n^{\alpha},S_j^{\beta}\}_{PB} = \delta_{ij}\epsilon_{\alpha\beta \gamma} S_j^{\gamma} , ~~~ \alpha,~\beta,~\gamma=1,2,3 ,  \ee where $\delta_{ij}$ is the Kronecker delta and $\epsilon_{\alpha\beta\gamma}$ is the Levi-Civita tensor, for any two functions ${\cal A}$ and ${\cal B}$ of spins one has  \be \{{\cal A},{\cal B}\}_{PB} =\sum_{\alpha,\beta,\gamma} \sum_{i=1}^{N}\epsilon_{\alpha\beta \gamma}\frac{\partial {\cal A}}{\partial S_n^{\alpha}}\frac{\partial {\cal B}}{\partial S_n^{\beta}} S_n^{\gamma} , \ee
and deduce the equation of motion for the spins of the lattice specified by the Hamiltonian (\ref{hamil}) as   
\begin{eqnarray} 
\label{vecform}
\frac{d\vec{S}_n}{dt}&=&\vec{S_n}\times[A(S_{n+1}^x+S_{n-1}^x)\vec{i} 
+B(S_{n+1}^y+S_{n-1}^y)\vec{j} + C(S_{n+1}^z+S_{n-1}^z)\vec{k} +2 D S_n^z\vec{k}]  
\nonumber \\ 
&+& \vec{S}_n\times\vec{\cal H},\;\;\;\; 
n=1,2,...,N.
\end{eqnarray}  
In Eq. (\ref{vecform})  $\vec{i},\,\vec{j},\,\vec{k}$ form a triad of Cartesian unit vectors. Rewriting Eq.  (\ref{vecform}) explicitly in component form, one obtains
\bea
\label{cartformx}
\frac{d{S_n^x}}{dt}&=&CS_n^y(S_{n+1}^z+S_{n-1}^z) 
-BS_n^z(S_{n+1}^y+S_{n-1}^y)-2DS_n^yS_n^z , \\ 
 \frac{d{S_n^y}}{dt}&=&AS_n^z(S_{n+1}^x+S_{n-1}^x) 
-CS_n^x(S_{n+1}^z+S_{n-1}^z)+2DS_n^xS_n^z + {\cal H}S_n^z,  \\
 \frac{d{S_n^z}}{dt}&=&BS_n^x(S_{n+1}^y+S_{n-1}^y) 
-AS_n^y(S_{n+1}^x+S_{n-1}^x) - {\cal H}S_n^y, ~n=1,2,...,N,   
\label{cartformz}
\eea
subject to the length constraint (\ref{constt}) on the spin vectors. To understand the underlying spin dynamics one has to analyze the initial value problem of the system of coupled nonlinear ordinary differential equations (\ref{vecform}) or (\ref{cartformx})-(\ref{cartformz}) along with the constraint (\ref{constt}) on the spin vectors, subject to appropriate boundary conditions like $\vec S_{n\rightarrow\infty}\rightarrow (\pm1,0,0)$ or $\vec S_{n \rightarrow\infty}\rightarrow(0,0,\pm1)$.  Although the system of differential equations (\ref{cartformx})-(\ref{cartformz}) is in general nonintegrable, it admits a few classes of exact solutions \cite{LS}. In this connection, the above type of spin evolutions are known to admit intrinsic localized modes (ILM), essentially through numerical analysis \cite{zolo,khalack02,sievers97,speight,nag}. Consequently, it will be of interest to look for exact solutions of ILM type by making appropriate ansatzes. In this paper, we demonstrate that explicit one-, two- and three-spin excitations can be identified in terms of Jacobian elliptic functions.

\section{Intrinsic Localized Modes: Exact Solutions} 
\label{ilm}
\subsection{One-spin excitation} 
\label{ilm1}
Let us consider the special set of solutions of the chain:  
\be
\label{1spin}
\vec S_n = ..., (1,0,0), (1,0,0), (S_i^x(t), S_i^y(t), S_i^z(t)), (1,0,0), (1,0,0), ... 
\ee 
(It may be noted that in the above we have used the notation suffix `$n$' to denote the general spin in the lattice, while the suffix `$i$' represents localized spin excitation). Then redesignating $(S_i^x(t), S_i^y(t), S_i^z(t))$ as $(S_0^x(t), S_0^y(t), S_0^z(t))$, the equation of motion for the excited spin at the lattice site $n$ reduces to 
\bea
\frac{dS_0^x}{dt} &=& -2DS_0^yS_0^z, \label{1spinx} \\ 
\frac{dS_0^y}{dt} &=& (2A+H)S_0^z + 2DS_0^xS_0^z, \label{1spiny} \\
\frac{dS_0^z}{dt} &=& -(2A+H)S_0^y .  
\label{1spinz}
\eea
Note that $\vec{S_n} = ..., (1,0,0), (1,0,0), (S_i^x(t), -S_i^y(t), -S_i^z(t)), (1,0,0), (1,0,0), ... $ is also a solution, if (\ref{1spin}) is a solution to (\ref{cartformx})-(\ref{cartformz}). The above solutions correspond to one spin excited exact localized solutions.

For $D=0$ the equations (\ref{1spinx})-(\ref{1spinz}) can be simplified and integrated easily to obtain the ILM solution given by trigonometric functions  
\be 
\label{1spinsol}
S_0^x=\sqrt{1-a^2}, ~~~ S_0^y=a\sin(\Omega t + \delta), ~~~ S_0^z=a\cos(\Omega t + \delta), ~~~\Omega=(2A+H),   
\ee 
where $a$ is an arbitrary constant and $\delta$ is a phase factor. 

For the case $D\ne0$, from Eq. (\ref{1spinz}) we get 
\be
\label{eqn00}
S_0^y = -\frac{1}{2A+H}\frac{dS_0^z}{dt}, 
\ee 
and from Eq. (9) we find 
\be 
\label{eqn01}
\frac{dS_0^y}{dt} = -\frac{1}{2A+H}\frac{d^2S_0^z}{dt^2} 
= [(2A+H) + 2DS_0^x)]S_0^z. \ee 
One can easily check that from (\ref{eqn01}) we can write  
\be 
\label{eqn02}
S_0^x = -\frac{1}{2A+H}\left[\frac{1}{S_0^z}\left(\frac{d^2S_0^z}{dt^2} 
\right) + (2A+H)^2\right]. \ee 
Substituting Eqs. (\ref{eqn00}) - (\ref{eqn02}) into Eq. (\ref{1spinx}) we find 
\be 
\label{eqn03}
\frac{dS_0^x}{dt} \equiv -\frac{1}{2D(2A+H)}\left[-\frac{1}{(S_0^z)^2} 
\left(\frac{dS_0^z}{dt}\right)\left(\frac{d^2S_0^z}{dt^2}\right) 
+\frac{1}{S_0^z}\frac{d^3S_0^z}{dt^3}\right] = \frac{2D}{(2A+H)}S_0^z 
\left(\frac{dS_0^z}{dt}\right) , 
\ee 
which implies 
\be
\label{eqn04}
\frac{1}{S_0^z}\frac{d^3S_0^z}{dt^3}  -\frac{1}{(S_0^z)^2} 
\left(\frac{dS_0^z}{dt}\right)\left(\frac{d^2S_0^z}{dt^2}\right) 
= -4D^2S_0^z\left(\frac{dS_0^z}{dt}\right) . 
\ee 

By letting $d^2S_0^z/dt^2=v(t)$, differentiating it, rearranging the 
previous equation and then dividing by $dS_0^z/dt$, we get 
\be 
\label{eqn05}
\frac{d}{dS_0^z}\left(\frac{v}{S_0^z}\right) = -4D^2S_0^z . 
\ee 
With $b$ an integration constant, we finally get  
\be 
\label{eqn06}
\frac{d^2S_0^z}{dt^2} + bS_0^z + 2D^2(S_0^z)^3 = 0.
\ee  
The solution of this equation can be easily written as \cite{grad,byrd,ml}
\be 
\label{eqn06}
S_0^z=a~\cn(\hat{\Omega}t+\delta,k),
\ee  
where $k$ is the modulus of the Jacobian elliptic function.

Thus, using (\ref{eqn06}) in (\ref{eqn00}) and (\ref{eqn02}), for $D\ne0$ the ILM solution is given by 
Jacobi elliptic functions,   
\be 
\label{eqn07}
S_0^x = \frac{Da^2}{2A+H}\cn^2(\hat{\Omega}t+\delta,k) + d , 
\ee 

\be 
\label{eqn08}
S_0^y = \frac{a\hat{\Omega}}{2A+H}[\sn(\hat{\Omega}t+\delta,k) 
\dn(\hat{\Omega}t+\delta,k)] , 
\ee 
\be
\label{eqn09} 
S_0^z = a~\cn(\hat{\Omega}t+\delta,k) , 
\ee 
where 
\be 
\label{eqn10} 
\hat{\Omega} = \sqrt{b+2a^2D^2} ,  ~~~ k^2=\frac{a^2D^2}{\hat{\Omega}^2}.   
\ee 

The arbitrary constants $b$ and $d$ in Eqs. (\ref{eqn07}) - (\ref{eqn10}) are determined from 
the constraint of unit spin length: $(S_i^x)^2+(S_i^y)^2+(S_i^z)^2=1.$ 
Inserting the explicit solutions for the three components of spin and matching 
coefficients of $\sn^4(\hat{\Omega}t+\delta,k)$, $\sn^2(\hat{\Omega}t+\delta,k)$ 
and constant terms, we find that    
\be 
\label{eqn11} 
d = \sqrt{1-a^2} - \frac{Da^2}{2A+H},  
\ee 
\be 
\label{eqn12}
b=(2A+H)^2-2a^2D^2 + 2D(2A+H)\sqrt{1-a^2}.  
\ee 
Note that the above solution (\ref{eqn07}) - (\ref{eqn09}) does exist in the anticontinuum limit $A = B = C = 0$ (see Eqs. (\ref{cartformx}) - (\ref{cartformz})). So the solution (\ref{eqn07}) - (\ref{eqn09}) may be considered as an analytic continuation of the anticontinuum limit in the discrete case.
The formation energy of an ILM is calculated by substituting the solution 
for the ILM in the Hamiltonian.  We find, after simplifying algebra, that  
\be 
\label{eqn13}
E = 2d(2A+H) = 2\sqrt{1-a^2}(2A+H) -2Da^2 + \{vacuum~state~energy - A\} . 
\ee 
The one spin excitation is schematically shown in Fig. \ref{spin1}. The above excitation can also be semiclassically quantized as shown in Sec. \ref{semiclass} below.
\begin{figure}
\includegraphics[width=1.0\columnwidth]{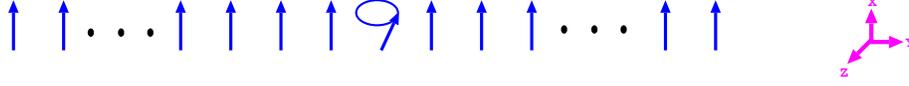}
\begin{center}
\caption{Schematic representation of one-spin excitation.}
\label{spin1}
\end{center}
\end{figure}

\subsection{Two-spin excitations} 
\label{ilm2}
We next consider localized two-spin excitations of the form 
\be
\label{eqn14} 
\vec{S}_n = ..., (1,0,0), (1,0,0), (S_i^x, S_i^y, S_i^z), (S_{i+1}^x, S_{i+1}^y, 
S_{i+1}^z), (1,0,0), (1,0,0), ...  
\ee 
\be
\label{eqn15}
= ..., (1,0,0), (1,0,0), (S_0^x, S_0^y, S_0^z), (S_{1}^x, S_{1}^y, 
S_{1}^z), (1,0,0), (1,0,0), ... 
\ee 
Equivalently one can also choose 
\be 
\label{eqn16}
\vec{S}_n = ..., (1,0,0), (1,0,0), (S_{i-1}^x, S_{i-1}^y, S_{i-1}^z), (S_{i}^x, S_{i}^y, 
S_{i}^z), (1,0,0), (1,0,0), ... 
\ee 
Then the defining equations for $\vec{S}_0(t)$ and $\vec{S}_1(t)$ become (from Eqs. 
(\ref{cartformx}) - (\ref{cartformz}) by using (\ref{eqn15})) 
\be 
\label{eqn17}
\frac{dS_0^x}{dt} = CS_0^yS_1^z - BS_0^zS_1^y - 2DS_0^yS_0^z , 
\ee 
\be 
\label{eqn18}
\frac{dS_0^y}{dt} = AS_0^z(S_1^x+1) - CS_0^xS_1^z + 2DS_0^xS_0^z + HS_0^z , 
\ee 
\be 
\label{eqn19}
\frac{dS_0^z}{dt} = BS_0^xS_1^y - AS_0^y(1+S_1^x) - HS_0^y , 
\ee 
\be 
\label{eqn20}
\frac{dS_1^x}{dt} = CS_0^zS_1^y - BS_1^zS_0^y - 2DS_1^yS_1^z , 
\ee 
\be 
\label{eqn21}
\frac{dS_1^y}{dt} = AS_1^z(S_0^x+1) - CS_1^xS_0^z + 2DS_1^xS_1^z + HS_1^z , 
\ee 
\be 
\label{eqn22}
\frac{dS_1^z}{dt} = BS_1^xS_0^y - AS_1^y(1+S_0^x) - HS_1^y . 
\ee

While the general solution of the system (\ref{eqn17}) - (\ref{eqn22}) is not known, a special class of 
solutions can be found with the choice 
\be 
\label{eqn23}
\vec{S}_0(t) = \vec{S}_1(t) . 
\ee 
In this case the system (\ref{eqn17}) - (\ref{eqn22}) degenerates into the following system of three 
coupled equations for any one of the vectors $\vec{S}_0(t)$ or $\vec{S}_1(t)$: 
\be 
\label{eqn24}
\frac{dS_0^x}{dt} = (C-B-2D)S_0^yS_0^z , 
\ee 
\be 
\label{eqn25}
\frac{dS_0^y}{dt} = (A+H)S_0^z + (A-C+2D)S_0^xS_0^z , 
\ee 
\be 
\label{eqn26}
\frac{dS_0^z}{dt} = (B-A)S_0^xS_0^y - (A+H)S_0^y . 
\ee
From (\ref{eqn24}) we can write 
\be 
\label{eqn27}
S_0^z = \frac{1}{(C-B-2D)S_0^y} \frac{dS_0^x}{dt} . 
\ee 
On the other hand from (\ref{eqn25})
\be 
\label{eqn28}
S_0^z = \frac{1}{[(A+H) + (A-C+2D)S_0^x]} \frac{dS_0^y}{dt} . 
\ee 
Equating (\ref{eqn27}) and (\ref{eqn28}), we obtain  
\be 
\label{eqn29}
\frac{1}{(C-B-2D)S_0^y} \frac{dS_0^x}{dt} = \frac{1}{[(A+H) + (A-C+2D)S_0^x]} \frac{dS_0^y}{dt} . 
\ee
Or equivalently 
\be 
\label{eqn30}
(C-B-2D)S_0^ydS_0^y = [(A+H) + (A-C+2D)S_0^x]dS_0^x . 
\ee 
Integrating, we have 
\be 
\label{eqn31}
(C-B-2D)\frac{(S_0^y)^2}{2} = [(A+H)S_0^x + (A-C+2D)\frac{(S_0^x)^2}{2} + \gamma] , 
\ee 
where $\gamma$ is an integration constant. 

Thus, we can reexpress $S_0^y$ and $S_0^z$ in terms of $S_0^x$ as follows. Firstly, from (\ref{eqn31}) we can write for the choice of parameters $C > (B-2D) > (A-2D) > 0$
\be 
\label{eqn32}
S_0^y=\sqrt{\frac{2}{(C-B-2D)}} \sqrt{C-A-2D} \left\{\frac{\gamma}{(C-A-2D)} + \frac{(A+H)}{(C-A-2D)}S_0^x - (S_0^x)^2 \right\}^{1/2}.  
\ee 
This can be simplified further as
\be 
\label{eqn33}
S_0^y = \sqrt{\frac{2(C-A-2D)}{(C-B-2D)}(a-S_0^x)(c+S_0^x) } , 
\ee 
where 
\be 
\label{eqn34}
a = \frac{1}{2} \left[\frac{(A+H)}{(C-A-2D)} + \sqrt{\frac{(A+H)^2}{(C-A-2D)^2} + \frac{4\gamma}{(C-A-2D)} }   \right] , 
\ee  
\be 
\label{eqn35}
c =\frac{1}{2} \left[- \frac{(A+H)}{(C-A-2D)} + \sqrt{\frac{(A+H)^2}{(C-A-2D)^2} + \frac{4\gamma}{(C-A-2D)} }   \right] , ~~~ a>c>0 . 
\ee 
Note that when $C>B-2D>A-2D>0$, the integration constant $\gamma$ has to be positive.  For other choices of the parameters $A,B,C,D$, similar decomposition can be given. 

Then from the relation $(S_0^x)^2+(S_0^y)^2 )+(S_0^z)^2 =1$, one can write
\be \small
\label{eqn36}
S_0^z = \left[1+\frac{C-A-2D)}{C-B-2D)}\right]^{\frac{1}{2}} \Bigg[\frac{(C-B-2D-2\gamma)}{(C-B-2D)+(C-A-2D)} \\
-  \frac{2(A+H)}{(C-B-2D)  + (C-A-2D)}S_0^x - (S_0^x)^2 \Bigg]^{\frac{1}{2}}  
\ee   
\be 
\label{eqn37}
= \left[1+\frac{C-A-2D}{C-B-2D}\right]^{1/2} [(b+S_0^x)(d-S_0^x)]^{1/2} , 
\ee 
where 
\be 
\label{eqn38}
b = \frac{(A+H) + \sqrt{(A+H)^2 + (C-B-2D-2\gamma)\{(C-B-2D)+(C-A-2D)\} } } {[(C-B-2D) + (C-A-2D)] } , 
\ee 
\be 
\label{eqn39}
d = \frac{-(A+H) + \sqrt{(A+H)^2 +(C-B-2D-2\gamma)\{(C-B-2D)+(C-A-2D)\} } } {[(C-B-2D) + (C-A-2D) ]} , 
\ee 
with $b>d>0$.   Here we restrict the arbitrary constant $\gamma$ such that $(C-B-2D-2\gamma) > 0$ so that $S_0^z$ is real in (49). 

Then using the expressions (\ref{eqn33}) and (\ref{eqn36}), we can rewrite Eq. (\ref{eqn24}) as 
\be \small
\label{eqn40}
\frac{dS_0^x}{\sqrt{(a-S_0^x)(b+S_0^x)(c+S_0^x)(d-S_0^x)}} = 
	[(C-B-2D)+(C-A-2D)]^{1/2}[2(C-A-2D)]^{1/2}dt , ~~ (a>b>c>d>0). 
\ee 
Integrating the above equation \cite{grad} and rearranging we can write down the solution as 
\be 
\label{eqn41}
S_0^x(t) = \frac{ \Gamma c ~\sn^2(\Omega t+\delta) -b} {1 - \Gamma \sn^2(\Omega t +\delta) } , ~~~ \Gamma < 1 , 
\ee
where the frequency 
\be 
\label{eqn42a}
\Omega = \sqrt{\frac{(a+b)(c+d)}{(C-A-2D)[(C-A-2D)+(C-B-2D)]} }, 
\ee 

\be 
\label{eqn42}
\Gamma = \frac{b+d}{c+d} , 
\ee 
%\be \label{eqn42a} \frac{dS_0^x}{dt} = (C-B-2D)(a-S_0^x)(b+S_0^x)(c+S_0^x)(d-S_0^x) \ee 
%\be \label{eqn43}  \Omega = \sqrt{\frac{(a+b)(c+d)}{(C-A-2D)\{(C-A-2D)+(C-B-2D)\} }  } \ee
and the square of the modulus of the Jacobi elliptic function  is
\be 
\label{eqn44}
k^2 = \sqrt{\frac{(b+d)(a+c)}{(a+b)(c+d)} } . 
\ee 

Knowing $S_0^x(t)$, we can write down the form of $S_0^y(t)$ and $S_0^z(t)$ straightforwardly from (\ref{eqn27}) and (\ref{eqn28}):
\be 
\label{eqn44a}
S_0^y(t) = \sqrt{\frac{ 2(b-c)(A-C+2D)(a+b-(a+c)\Gamma\sn^2(\Omega t)} {(C-B-2D)(1 - \Gamma \sn^2(\Omega t ))^2 }} , 
\ee
and
\be
\label{eqn44b}
S_0^z(t) = \sqrt{1-{(S_0^x)}^2 - {(S_0^y)}^2}.
%\frac{2(b-c)\Gamma \Omega\sn(\Omega t )\cn(\Omega t )\dn(\Omega t )  }{(1 - \Gamma \sn^2(\Omega t ))^2) \left[A+H+(A-C+2D)\sqrt{\frac{ 2(b-c)(A-C+2D)(a+b-(a+c)\Gamma\sn^2(\Omega t)} {(C-B-2D)(1 - \Gamma \sn^2(\Omega t )^2) }} }\right], 
\ee
 The corresponding energy values can be calculated from the form of the Hamiltonian (\ref{hamil}).  

A similar analysis can be carried out for the case $A > B > C$. Special cases of equality in signs 
(easy axis and easy plane) can be easily dealt with in the above analysis.  

\begin{figure}
\includegraphics[width=1.0\columnwidth]{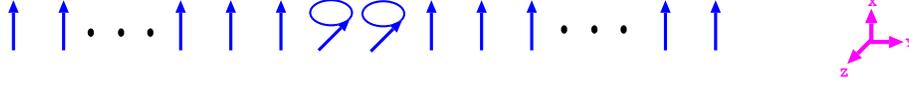}
\begin{center}
\caption{Schematic representation of two-spin excitation (\ref{eqn41})-(\ref{eqn44b}).}
\label{spin2a}
\end{center}
\end{figure}

\begin{figure}
\includegraphics[width=1.0\columnwidth]{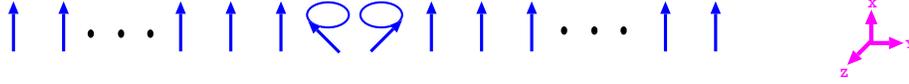}
\begin{center}
\caption{Schematic representation of two-spin excitation (\ref{eqn45}).}
\label{spin2b}
\end{center}
\end{figure}

Finally we also note that the system (\ref{eqn17}) - (\ref{eqn22}) also admits a solution 
\be 
\label{eqn45}
\vec{S}_1(t) = (S_0^x, -S_0^y, -S_0^z) , 
\ee 
if $\vec{S}_1(t) = \vec{S}_0(t)$ is a solution.  It corresponds to the motion of the two neighboring 
spins in the opposite directions in the y-z plane.  Note that the corresponding energy value is 
different from that of the excitation $\vec{S}_1(t) = \vec{S}_0(t)$. The above two types of solutions are shown schematically in Figs. \ref{spin2a} and \ref{spin2b}.

In the above we have used only Eqs. (\ref{eqn24}) and (\ref{eqn26}) and not (\ref{eqn25}).  However, this equation is automatically satisfied because $S_i^x(dS_i^x/dt) + S_i^y(dS_i^y/dt) + S_i^z 
(dS_i^z/dt) = 0$ due to the constraint ${\vec{S}_i}^2 = 1$. We also note here that the above solutions do exist in the anticontinuum limit $A=B=C=0$, by choosing the integration constant suitably or choosing the sign of $D$.

\subsection{Three-spin excitations} 
\label{ilm3}
We now look for a solution of the type 
\be 
\label{eqn46}
\vec{S}_n(t) = ..., (1,0,0), (1,0,0), (S_{-1}^x, S_{-1}^y, S_{-1}^z), (S_0^x, S_0^y, S_0^z),(S_{1}^x, S_{1}^y, S_{1}^z), (1,0,0), (1,0,0), ... 
\ee 
This is a three spin excited localized solution.
Then from (\ref{cartformx}) - (\ref{cartformz}), we can write 
\be 
\label{eqn47}
\frac{dS_{-1}^x}{dt} = CS_{-1}^yS_0^z -BS_{-1}^zS_0^y - 2DS_{-1}^yS_{-1}^x , 
\ee 
\be 
\label{eqn48}
\frac{dS_{-1}^y}{dt} = AS_{-1}^z(1+S_0^x)  -CS_{-1}^xS_0^z + 2DS_{-1}^xS_{-1}^z + HS_{-1}^z , 
\ee 
\be
\label{eqn49} 
\frac{dS_{-1}^z}{dt} = BS_{-1}^xS_0^y  -AS_{-1}^y(1+S_0^x) - HS_{-1}^y , 
\ee  
\be \label{eqn50} 
\frac{dS_0^x}{dt} = CS_0^y(S_1^z+S_{-1}^z) -BS_0^z(S_1^y+S_{-1}^y) - 2DS_0^yS_0^x , 
\ee 
\be \label{eqn51} 
\frac{dS_0^y}{dt} = AS_0^z(S_1^x+S_{-1}^y) -CS_0^x(S_1^z+S_{-1}^z) +2DS_0^xS_0^z + HS_0^z, 
\ee 
\be \label{eqn52} 
\frac{dS_0^z}{dt} = BS_0^x(S_1^y+S_{-1}^y) - AS_0^y(S_1^x+S_{-1}^x) -  HS_0^y , 
\ee 
\be \label{eqn53} 
\frac{dS_1^x}{dt} = CS_1^yS_0^z - BS_1^zS_0^y - 2DS_1^yS_1^x , 
\ee 
\be \label{eqn54} 
\frac{dS_1^y}{dt}  = AS_1^z(1+S_0^x) - CS_1^xS_0^z + 2DS_1^xS_1^z + HS_1^z , 
\ee 
\be \label{eqn55} 
\frac{dS_1^z}{dt} = BS_1^xS_0^y - AS_1^y(S_0^x+1) - HS_1^y . 
\ee 

One can easily check that $\vec{S_{1}}(t) = \vec{S_0}(t) = \vec{S_{-1}}(t)$ is not an allowed solution.  However, if one chooses the middle spin to be in the ground state,
\be \label{eqn56} 
\vec{S}_0(t) = (1,0,0) , 
\ee 
while the two neighbors $\vec{S}_{-1}(t)$ and $\vec{S}_1(t)$ evolve in time, the above set of equations (\ref{eqn47}-\ref{eqn55}) reduces to 
\be \label{eqn57} 
\frac{dS_{-1}^x}{dt} = -2DS_{-1}^yS_{-1}^z , 
\ee 
\be \label{eqn58} 
\frac{dS_{-1}^y}{dt} = (2A+H)S_{-1}^z + 2DS_{-1}^xS_{-1}^z , 
\ee 
\be \label{eqn59} 
\frac{dS_{-1}^z}{dt} = -(2A+H)S_{-1}^y , 
\ee 
\be \label{eqn60} 
0 = 0 
\ee 
\be \label{eqn61} 
0 = -C(S_{1}^z + S_{-1}^z) , 
\ee 
\be \label{eqn62} 
0 = B(S_1^y + S_{-1}^y) , 
\ee 
\be \label{eqn63} 
\frac{dS_1^x}{dt} = -2DS_1^yS_1^x , 
\ee 
\be \label{eqn64} 
\frac{dS_1^y}{dt} = (2A+H)S_1^z + 2DS_1^xS_1^z , 
\ee 
\be \label{eqn65} 
\frac{dS_1^z}{dt} = -(2A+H)S_1^z . 
\ee 

\begin{figure}
\includegraphics[width=1.0\columnwidth]{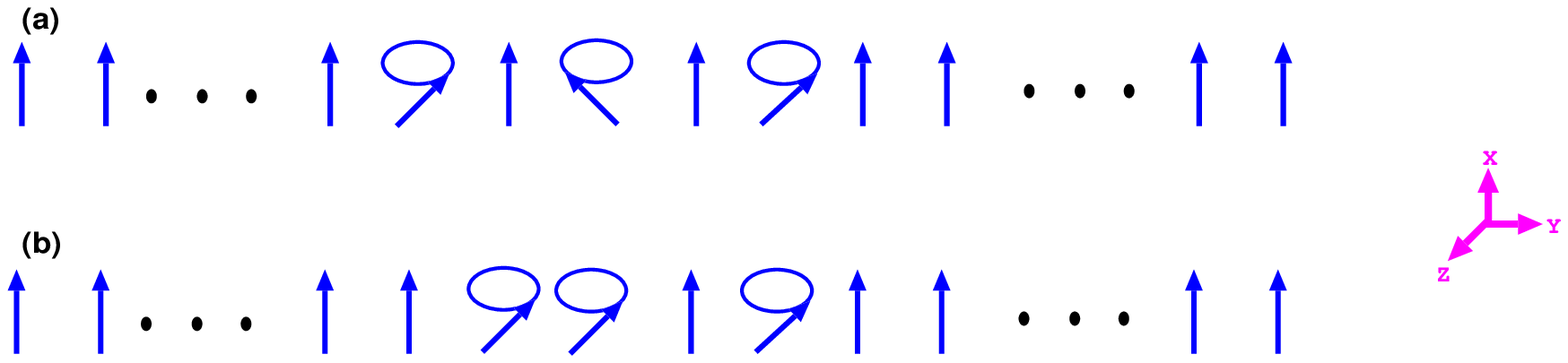}
\begin{center}
\caption{Schematic representation of three-spin excitation given by (a) Eq. (\ref{eqn3sa}) and (b) Eq. (\ref{eqn3sb}).}
\label{spin3a}
\end{center}
\end{figure}

Eqs. (\ref{eqn61}) - (\ref{eqn62}) simplify as
\be \label{eqn66} 
S_1^z = -S_{-1}^z , 
\ee 
\be \label{eqn67} 
S_1^y = -S_{-1}^y . 
\ee 
Consequently if $\vec{S}_{-1}(t) = (S_{-1}^x, S_{-1}^y, S_{-1}^z)$ is a solution of (\ref{eqn57}) - (\ref{eqn59}), 
then 
\be \label{eqn68} 
\hat{\vec{S}}_1(t) = (S_{-1}^x, -S_{-1}^y, -S_{-1}^z) 
\ee 
is indeed a solution compatible with (\ref{eqn63}) - (\ref{eqn65}).  Thus the three-spin excitation case 
degenerates here into the two isolated spin excitations, 
\be \label{eqn69} 
\vec{S}_n(t) = ..., (1,0,0), (1,0,0), (S_{-1}^x, S_{-1}^y, S_{-1}^z), (1,0,0), (S_{-1}^x, -S_{-1}^y, -S_{-1}^z), (1,0,0), (1,0,0), ... 
\ee 
\begin{figure}
\includegraphics[width=1.0\columnwidth]{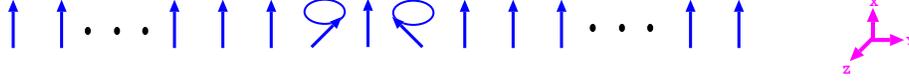}
\begin{center}
\caption{Schematic representation of another two-spin excitation given by (\ref{eqn69}).}
\label{spin3}
\end{center}
\end{figure}
Note that the form of $\vec{S}_{-1}(t)$ is exactly the one-spin excitation expression given 
by (\ref{eqn06}) - (\ref{eqn08}), since the form of (\ref{eqn47}) - (\ref{eqn49}) is the same as that of (\ref{1spinx}) - (\ref{1spinz}).  It is easy to check that the energy of this excitation is twice that of the single spin excitation. Obviously the above two-spin excitation also has a smooth anticontinuum limit. Solution (\ref{eqn69}) is also shown schematically in Fig. \ref{spin3}. Now one can generalize the above analysis to higher spin excitations and following the analysis given above, one can obtain a three spin excitation of the following forms:
%\begin{itemize}
%\item 
\be 
\label{eqn3sa}
(a)~~~\vec{S}_n(t) = ..., (1,0,0), (S_{-1}^x, S_{-1}^y, S_{-1}^z), (1,0,0), (S_{-1}^x, -S_{-1}^y, -S_{-1}^z), (1,0,0), (S_{-1}^x, S_{-1}^y, S_{-1}^z), (1,0,0), ...
\ee
%\item 
\be
\label{eqn3sb}
(b)~~~~\vec{S}_n(t) = ..., (1,0,0), (S_{0}^x, S_{0}^y, S_{0}^z), (S_{0}^x, S_{0}^y, S_{0}^z), (1,0,0), (S_{-1}^x, S_{-1}^y, S_{-1}^z), (1,0,0) ...
\ee
%\end{itemize}
and various other combinations. The three-spin excitations (\ref{eqn3sa}) and (\ref{eqn3sb}) are shown in Fig. \ref{spin3a}. 
The analysis can be generalized to higher spin excitations like four-spin and five-spin 
excitations.  One can obtain analogous results.  These will be reported separately elsewhere. 

\section{Linear stability} 
\label{stability}
The solutions reported in the earlier section, though exact, are all particular or special solutions of 
the full dynamical equations.  It is then important to analyze their stability, at least their  
linear stability for small perturbations.  For this purpose, we perturb the solutions 
for the spins as 
\be \label{eqn70} 
\vec{S}'_n(t) = \vec{S}_n(t) + \epsilon\vec{\rho}_n(t) , ~~~ \epsilon\ll 1 
\ee 
in Eqs. (\ref{vecform}) and retain only the terms linear in $\epsilon$.  Due to the constancy of the 
length of the spins, we also have the constraints 
\be \label{eqn71} 
\vec{S}_n\cdot\vec{\rho}_n = 0 . 
\ee
Consequently, the linearized equations can be written down as 
\be \small \label{eqn72} 
\frac{d\rho_n^x}{dt} = C\rho_n^y(S_{n+1}^z + S_{n-1}^z) + CS_n^y(\rho_{n+1}^z+\rho_{n-1}^z) 
-B\rho_n^z(S_{n+1}^y+S_{n-1}^y) -BS_n^z(\rho_{n+1}^y+\rho_{n-1}^y) -2D(\rho_n^yS_n^z+S_n^y\rho_n^z) , 
\ee 
\be \small\label{eqn73} 
\frac{d\rho_n^y}{dt} = A\rho_n^z(S_{n+1}^x + S_{n-1}^x) + AS_n^z(\rho_{n+1}^x+\rho_{n-1}^x) 
-C\rho_n^x(S_{n+1}^z+S_{n-1}^z) -CS_n^x(\rho_{n+1}^x+\rho_{n-1}^x) +2D(\rho_n^zS_n^x+S_n^z\rho_n^x) - H\rho_n^z,  
\ee 
\be \small\label{eqn74} 
\frac{d\rho_n^z}{dt} = B\rho_n^x(S_{n+1}^y + S_{n-1}^y) + BS_n^x(\rho_{n+1}^y+\rho_{n-1}^y) 
-AS_n^y(\rho_{n+1}^x+S_{n-1}^x) -A\rho_n^y(S_{n+1}^x+S_{n-1}^x) -H\rho_n^y ,  
\ee 
with the constraint 
\be \label{eqn75} 
S_n^x\rho_n^x + S_n^y\rho_n^y + S_n^z\rho_n^z =  0 . 
\ee 
\begin{figure}
\begin{center}
\includegraphics[width=0.7\columnwidth]{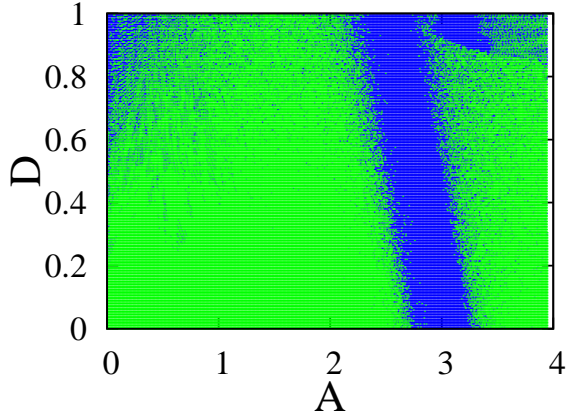}
\caption{(color online) Stability of single spin excitation (\ref{eqn07}) - (\ref{eqn10}) in the ($A,D$) parameter space with other parameters chosen as $B=0.2, C=0.1, H=0.5$ and arbitrary constant as $a=0.8$, for $N=21$ spins. Here the green regimes are stable regions, while the blue regimes are unstable where the Floquet multiplier $|\lambda_i| > 1$.}
\label{stab}
\end{center}
\end{figure}
For each of the exact solutions discussed in the previous section, we can replace $\vec{S}_n(t)$ 
in (\ref{eqn63}) - (\ref{eqn65}) by the appropriate form and then consider the linearized equation.  For example, for the one-spin excitation, we replace $\vec{S}_n(t)$ by the form (\ref{1spin}) with $\vec{S}_i(t)$ given by Eqs. (\ref{eqn07}) - (\ref{eqn09}).  The resultant Floquet equation with periodic functions can be analyzed 
for Floquet exponents numerically and the region of stability in the parameter space can be 
identified.  As in the case of several systems studied in the literature, one can expect small 
amplitude excitations to be linearly stable.  Equivalently the problem can be expressed in 
terms of stereographic coordinates as has been done by Zolotaryuk et al. \cite{zolo} to find the 
Floquet matrix numerically. In Fig. \ref{stab}, we have presented the details of the linear stability analysis for the single spin excitation given by the expressions (\ref{eqn07}) - (\ref{eqn10}) by evaluating the associated Floquet multipliers. The green regions correspond to linearly stable excitations where all the Floquet multipliers $|\lambda_i| \le 1$, while the unstable regions are represented by blue color in the ($A-D$) parameter space. 

\section{Semiclassical quantization of ILMs} 
\label{semiclass}
Due to the periodic nature of the exact localized solutions obtained in the manuscript one can carry out a semiclassical quantization of the spin excitations using Einstein-Brillouin-Keller (EBK) or Bohr-Sommerfeld quantization rule as suggested by Aubry \cite{aubry97}.  Considering now the one-spin excitations, with the form (\ref{eqn07}) - (\ref{eqn09}), one can carry out a Bohr-Sommerfeld quantization 
\be \label{eqn76} 
\oint pdq = \left(n+\frac{1}{2}\right)h , ~~~ n=0,1,2, ... 
\ee 
where the canonically conjugate variables are chosen from (23) - (25) as 
\be \label{eqn77} 
p = S_0^z = a ~\cn(u,k) , ~~~ u = (\Omega t+\delta) 
\ee 
and 
\be \label{eqn78} 
q = \arctan(S_0^y/S_0^x) = \arctan\left(\frac{1}{k}\frac{[\sn(u,k)\dn(u,k)]}{[\eta+\cn^2(u,k)]}\right),
\ee 
where $\eta=\frac{d(2A+H)}{Da^2} = \frac{\sqrt{1-a^2}(2A+H)}{Da^2}-1$, using $d$ in Eq. (\ref{eqn11}).
Then the integral (\ref{eqn76}) can be written as 
\be \label{eqn79} 
\Omega \oint p\left(\frac{dq}{du}\right)du = \Omega a k \int_0^{4K}{\left[ \left( 1- \frac{\beta^2}{\alpha^2}\right) - \sn^2(u,k) + \frac{\beta^2\left(1+\frac{1}{\alpha^2}\right)}{1-\alpha^2\sn^2(u,k)} \right]du},
\ee
where  $\beta^2= \left[\frac{1-k^2(\eta+1)}{(k^2+1)(\eta+1)}\right] $, $\alpha^2= \left[\frac{2k^2(\eta+1)-1}{(k^2+1)(\eta^2+1)^2}\right]  $.  Carrying out the integrals on the right hand side \cite{byrd}, we finally obtain 

\be \label {eqn80}
{
 \hat{\Omega} a k \left[1 - \frac{A}{\alpha^2}K(k) 
- \frac{[K(k)-E(k)]}{k^2} + A\left(1+\frac{1}{\alpha^2}\right)\Pi(\alpha^2,k)\right]= \left(n+\frac{1}{2}h\right), ~~n=0,1,2,.. }
\ee 

%\frac{[K(k)-E(k)]}{k^2}   $u = (\hat{\Omega} t+\delta)$,
Here $K(k)$, $E(k)$ and $\Pi(\alpha^2,k)$ are the complete elliptic integrals of the first, second and third kinds, respectively \cite{grad}. Here we have used the formulas \cite{byrd}:
\be
\int{\sn^2(u)} du = \frac{1}{k^2}[u-E(u)],~~~~  [E(u)=E(\phi,k)] 
\ee and
\be
\int{\frac{du}{1-\alpha^2 \sn^2(u)}} du =  \Pi(\phi,\alpha^2,k) 
\ee
where  $E(\phi,k) $ and $\Pi(\phi,\alpha^2,k) $ are the incomplete elliptic integrals of the second and third kind, respectively.

Solving the above transcendental equation one can obtain the quantized values of $a$. Using the quantized values of the amplitude $a$ of oscillation, and substituting the same in the energy expression (\ref{eqn13}), we obtain the quantized energy levels for the one-spin excitation. Similar analysis can be carried out for the other spin excitations also using the explicit solutions given in the earlier sections.

\section{Summary and Conclusions} 
\label{conclusion}
In this paper we have shown that explicit intrinsic localized modes, in particular one-spin, two-spin and three-spin excitations, can be deduced analytically in terms of Jacobi elliptic functions for the anisotropic Heisenberg ferromagnetic spin chain in the presence of an external magnetic field. We also obtained the corresponding creation energy of these excitations. These solutions have proper anticontinuum limit and thus can be considered as analytic continuation of the former in the discrete cases. One can extend in principle to identify M-spin excitations ($M > 3$) as well. The stability regimes can be identified by considering the associated linear eigenvalue problem. We also obtained the semi-classical quantization of the ILMs. 

The different one-spin, two-spin and three-spin exact localized solutions we have reported in this paper correspond to different energies and can be observed in materials such as CsNiF$_3$ and TMMC, which represent examples of classical Heisenberg spins, under appropriate excitation energies.   The localization extent of each solution can be possibly inferred from magnon spectrum measurements.  An ability to controllably excite different ILMs in spin chains may lead to interesting spin-torque and related spintronic applications. We hope that the present results will give a motivation to search for more general explicit ILM solutions in magnetic and other lattices as well. It would be worthwhile to observe these spin excitations in magnon dispersion anomalies and directly using nonlinear spectroscopic techniques \cite{sato04}.

\section{Acknowledgments} 
The work is supported by the Department of Science and Technology (DST)--Ramanna program (ML), DST--IRHPA research project (ML and BS) and DAE Raja Ramanna Fellowship program (ML). ML acknowledges the hospitality of the Center for Nonlinear Studies at LANL where this work was initiated. This work was supported in part by the U.S.  Department of Energy.

\section*{References}
%\begin{references}

%\end{references} 

\end{document}